\documentclass[pra,twocolumn,showpacs,preprintnumbers,amsmath,amssymb,superscriptaddress]{revtex4}

\usepackage{graphicx} 

\newcommand{\bra}[1]{\langle#1|} 
\newcommand{\ket}[1]{|#1\rangle} 

\begin{document}

\title{Single observable concurrence measurement without simultaneous copies}

\author{A. Salles}
\email{salles@if.ufrj.br} \affiliation{Instituto de F\'{\i}sica, Universidade
  Federal do Rio de Janeiro, \\Caixa Postal 68528, Rio de Janeiro, RJ
  21941-972, Brazil}
\author{F. de Melo} \email{fmelo@if.ufrj.br} \affiliation{Instituto de
  F\'{\i}sica, Universidade Federal do Rio de Janeiro, \\Caixa Postal 68528,
  Rio de Janeiro, RJ 21941-972, Brazil} 
\author{J. C. Retamal}
\affiliation{Departamento de F\'{i}sica, Universidad de Santiago de Chile, Casilla 307,
Correo 2, Santiago, Chile}
\author{R. L. de Matos Filho}
\affiliation{Instituto de F\'{\i}sica, Universidade Federal do Rio de Janeiro,
  \\Caixa Postal 68528, Rio de Janeiro, RJ 21941-972, Brazil}
\author{N. Zagury}
\affiliation{Instituto de F\'{\i}sica, Universidade Federal do Rio de Janeiro,
  \\Caixa Postal 68528, Rio de Janeiro, RJ 21941-972, Brazil}

\date{\today}

%
%

\begin{abstract}
  We present a protocol that allows us to obtain the concurrence of any two
  qubit pure state by performing a minimal and optimal tomography of one of the subsystems
  through measuring a single observable of an ancillary four dimensional
  qudit. An implementation for a system of trapped ions is also proposed, which
  can be achieved with present day experimental techniques.
\end{abstract}

 \maketitle

%
%

Even if entanglement was spotted as a key feature of quantum mechanics since
the early days of the theory~\cite{schrodinger}, it was only with
the advent of quantum information science that a great deal of attention was
drawn upon the problems of characterizing, properly quantifying, and
ultimately measuring entanglement~\cite{bennett:247}.

Until recently, measurements of entanglement had only been achieved indirectly by
performing measurements on several non-commuting observables of the system, and then
adequately combining the results~\cite{james:052312}. A direct
measurement of concurrence (previously shown to be a proper
entanglement measure~\cite{wootters:2245}), however, was reported
in~\cite{walborn06}, in which two copies of the state were used
simultaneously, following the idea in~\cite{florian}. Although this
experiment constitutes a landmark on the path towards fully understanding
quantum entanglement, a simpler measurement, which does not involve simultaneous copies, is
desirable.

In the simplest case where we deal with a pure state of two qubits, one way to
address the problem is to take advantage of a well-known relation between the
bipartite concurrence and the reduced density matrix of either subsystem~\cite{coffman:052306}:
\begin{equation}
C^2=4 \det \rho_{q},
\label{concu}
\end{equation}
where $\rho_{q}$ is the reduced density matrix of one of the qubits and $C$ is the
bipartite concurrence. Hence, the concurrence of the system can be obtained by performing the tomography of only one of the qubits.

In ref.~\cite{rehacek:052321}, \v{R}eh\'{a}\v{c}ek~\emph{et al.} presented a protocol for optimal minimal qubit tomography, in which all information pertaining to the state of one qubit is
obtained by measuring the population of the states of two ancillary qubits,
which are previously entangled with the target qubit by nonlocal
operations.  In a loose sense, information of the measured qubit is
``written'' in the ancillas: the three values $\langle\hat{\sigma}_{x}\rangle $, $\langle\hat{\sigma}_{y}\rangle$ and $\langle\hat{\sigma}_{z}\rangle$ necessary for
qubit tomography are encoded into the four probabilities $P_{jk}$ $(j,k=0,1)$ of the ancillary system to be found in each state $\ket{jk}$, of which only three are independent because of the unity sum requirement.  An implementation where the qubit was encoded in linear photon polarizations and different paths in an optical interferometer played the role of ancillary qubits was also proposed. Later on, this proposal was experimentally achieved~\cite{ling:022309}.

The aim of this article is twofold: first, we want to point out the fact that
concurrence for a two qubit pure state can be obtained through measurement of
the probability distribution of the spectrum of a single observable, without the need of
simultaneous copies of the state. This is achieved by performing a minimal and optimal tomography of one of the qubits via
a single measurement on an ancillary four dimensional system. The tomography is minimal in the sense that no redundant information is obtained from the measurements (as opposed to the standard procedure) and optimal in that it achieves maximum accuracy in determining an unknown state~\cite{rehacek:052321}. Even if the procedure developed in~\cite{rehacek:052321} could in
principle be used for such an end, we will use in order to illustrate
our point a tomographic protocol of our own, in which the fact that the
probability distribution of the spectrum of a single
observable is being measured appears naturally. By introducing a four level qudit as our ancillary system, we are able to ``write'' the three desired values in the populations of the four levels of
the ancilla $P_{G}$, $P_{G'}$, $P_{E}$ and $P_{E'}$, of which only three are
independent. By choosing these states to have different energies, we can pick the observable to be the energy of the ancilla.

Second, we propose an implementation of the protocol for a system of trapped
 ions, which is achievable with present day experimental techniques. Our
 protocol proves simpler to implement for this kind of systems than that
 in~\cite{rehacek:052321}, for it uses one less ancilla, which means that one less
 ion is involved. 

In what follows, we will denote by $\ket{\chi}$ the two qubit pure
state on which tomography of one qubit is to be performed, and
$\ket{G}$, $\ket{G'}$, $\ket{E}$ and $\ket{E'}$ the four distinct
states of the ancilla. We will make use of two kinds of operations:

i) Rotations between ancillary states. These are denoted by
$\hat{R}_{\alpha}^{JK}(\theta )=\exp\left(-i\frac{\theta
  }{2}\hat{\sigma}_{\alpha }^{JK}\right)$, where $\hat{\sigma}_{\alpha }^{JK}$ is one of the Pauli operators
($\alpha\in\{x,y,z\}$) defined on the subspace spanned by the arbitrary states $\ket{J}$ and
$\ket{K}$:
\begin{equation}
\label{subspace sigmas}
\begin{split}
\hat{\sigma}_{x}^{JK} &=\ket{K}\bra{J}+\ket{J}\bra{K}\\
\hat{\sigma}_{y}^{JK} &=-i(\ket{K}\bra{J}-\ket{J}\bra{K})\\
\hat{\sigma}_{z}^{JK} &=\ket{K}\bra{K}-\ket{J}\bra{J}.\\
\end{split}
\end{equation}

ii) Controlled operations applied on the target qubit and controlled by the
ancilla.  These are denoted by $C^{A}\hat{U}$, where $A\in\{G, G', E, E'\}$ denotes
the control ancillary state whose occupation implies action of operator $\hat{U}$ on the selected
qubit. We can have, for instance: $C^{E}\hat{U}\frac{1}{\sqrt{2}}\left(\ket{G}+\ket{E}\right)\ket{\chi}=\frac{1}{\sqrt{2}}\left(\ket{G}+\ket{E}\hat{U}\right)\ket{\chi}$, where we explicitly put operator $\hat{U}$, which acts
only on the target qubit, to the right of the ancilla ket. Only three instances of $\hat{U}$ will be actually realized: $\hat{\sigma}_{x}$, $\hat{\sigma}_{y}$ and $-\hat{\sigma}_{z}$, which are unitary operations on the qubit, whose basis states are denoted by $\ket{g_{q}}$ and $\ket{e_{q}}$.

%
%

{\em Protocol.} Our protocol starts with the system in the state $\ket{G}\ket{\chi}.$ To this initial state, we apply three successive rotations, 
$R_y ^{GE}(\theta_1)$,  $R_y ^{GG'}(\theta_2)$ and $R_y ^{G'E'}(\theta_3)$ with $\theta_1$, $\theta_2$ and $\theta_3$ such that
\begin{equation}
\begin{split}
R_y ^{GE}(\theta_1)\ket{G} & = \frac{1}{\sqrt{6}}\left(\sqrt{5}\ket{G}+\ket{E}\right),  \\
R_y ^{GG'}(\theta_2)\ket{G} & = \frac{1}{\sqrt{5}}\left(\sqrt{3}\ket{G}+\sqrt{2}\ket{G'}\right), \\
R_y ^{G'E'}(\theta_3)\ket{G'} & =\frac{1}{\sqrt{2}}\left(\ket{G'}-\ket{E'}\right).
\end{split}
\end{equation} 

We thus obtain the state
\begin{equation}
\frac{1}{\sqrt{2}}\left(\ket{G}+\frac{1}{\sqrt{3}}\ket{G'}+\frac{1}{\sqrt{3}}\ket{E}-\frac{1}{\sqrt{3}}\ket{E'}\right)\ket{\chi}.
\label{preparedancilla}
\end{equation}

The next step of the protocol requires us to perform the controlled operations $C^{G'}(\hat{\sigma}_y)$, $C^{E}(\hat{\sigma}_x)$ and $C^{E'}(-\hat{\sigma}_z)$, ending up with the following state:
\begin{equation}
\frac{1}{\sqrt{2}}\left(\ket{G}+\frac{1}{\sqrt{3}}\ket{G'}\hat{\sigma}_y+\frac{1}{\sqrt{3}}\ket{E}\hat{\sigma}_x+\frac{1}{\sqrt{3}}\ket{E'}\hat{\sigma}_z\right)\ket{\chi}.
\label{aftercontrol}
\end{equation}

Finally we apply the following local $\pi/2$ rotations around the $y$ axis on the ancilla:
$\hat{R}_{y}^{GE}\left(\frac{\pi}{2}\right)$,
$\hat{R}_{y}^{G'E'}\left(\frac{\pi}{2}\right)$,
$\hat{R}_{y}^{GG'}\left(\frac{\pi}{2}\right)$ and $\hat{R}_{y}^{EE'}\left(\frac{\pi}{2}\right)$  to obtain:
\begin{equation}
\left(\ket{G}\hat{Q}_{G}-\ket{G'}\hat{Q}_{G'}-\ket{E}\hat{Q}_{E}+\ket{E'}\hat{Q}_{E'}\right)\ket{\chi},
\label{finalstate}
\end{equation}

where we have:
\begin{equation}
\begin{split}
\hat{Q}_{G} & =\frac{1}{2\sqrt{2}}\left( \hat{\openone}+\frac{1}{\sqrt{3}}\left(\hat{\sigma}_{x}+\hat{\sigma}_{y}+\hat{\sigma}_{z}\right)\right), \\
\hat{Q}_{G'} & =\frac{1}{2\sqrt{2}}\left( \hat{\openone}+\frac{1}{\sqrt{3}}\left(\hat{\sigma}_{x}-\hat{\sigma}_{y}-\hat{\sigma}_{z}\right)\right), \\
\hat{Q}_{E} & =\frac{1}{2\sqrt{2}}\left( \hat{\openone}+\frac{1}{\sqrt{3}}\left(-\hat{\sigma}_{x}+\hat{\sigma}_{y}-\hat{\sigma}_{z}\right)\right), \\
\hat{Q}_{E'} & =\frac{1}{2\sqrt{2}}\left( \hat{\openone}+\frac{1}{\sqrt{3}}\left(-\hat{\sigma}_{x}-\hat{\sigma}_{y}+\hat{\sigma}_{z}\right)\right).
\end{split}
\end{equation}

We then readily calculate the probabilities for the ancilla to be found in each state:
\begin{equation}
\label{probabilities}
\begin{split}
P_{G} & =\frac{1}{4}\left(1+\frac{1}{\sqrt{3}}(\langle\hat{\sigma}_{x}\rangle+\langle\hat{\sigma}_{y}\rangle+\langle\hat{\sigma}_{z}\rangle )\right) \\
P_{G'} & =\frac{1}{4}\left(1+\frac{1}{\sqrt{3}}(\langle\hat{\sigma}_{x}\rangle-\langle\hat{\sigma}_{y}\rangle-\langle\hat{\sigma}_{z}\rangle )\right) \\
P_{E} & =\frac{1}{4}\left(1+\frac{1}{\sqrt{3}}(-\langle\hat{\sigma}_{x}\rangle+\langle\hat{\sigma}_{y}\rangle-\langle\hat{\sigma}_{z}\rangle )\right) \\
P_{E'} & =\frac{1}{4}\left(1+\frac{1}{\sqrt{3}}(-\langle\hat{\sigma}_{x}\rangle-\langle\hat{\sigma}_{y}\rangle+\langle\hat{\sigma}_{z}\rangle )\right).
\end{split}
\end{equation}

By adding and subtracting these probabilities, we can obtain any mean value
$\langle\hat{\sigma}_{\alpha}\rangle$, which means we have successfully
performed the tomography of the qubit just by measuring the populations of the energy
eigenstates of the ancilla.

We would like to stress that the four values of the probabilities in Eq.~\ref{probabilities} are the
expectation values of the operators $\hat{Q}_{G}^2,$ $\hat{Q}_{G'}^2,$
$\hat{Q}_{E}^2$ and $\hat{Q}_{E'}^2$, which constitute a minimal and
optimal POVM~\cite{rehacek:052321} for the one qubit tomography we are
considering.


Having performed the tomography of one of the qubits, it is straightforward to find the value of
the concurrence of the bipartite system using relation~\eqref{concu}. In terms
of the occupation probabilities, it is
given by:
\begin{equation}
C^2=4\left(1-3(P_{G}^2+P_{G'}^2+P_{E}^2+P_{E'}^2)\right)
\end{equation}

We thus managed to obtain the value of the concurrence by measuring the
probability distribution of the spectrum of a single observable of a four dimensional ancillary system, with no need of simultaneous copies of the state. We stress the fact that even if only the expectation value of an observable is needed instead of the complete probability distribution in schemes using simultaneous copies, in practice this distribution must be determined anyway in order to compute the expectation value~\cite{walborn06}.

%
%

{\em Trapped Ions Implementation}. Consider now a system of ions inside a
linear Paul trap. To a good approximation, the effect of the trap in the
motion of the ions can be described by a harmonic oscillator.  The qubits are encoded in ground and excited electronic states of each ion, while one of the ions plays the role of the ancillary system. When an ion is illuminated
by laser light quasi-resonant with one of its electronic transitions, the collective motional
degrees of freedom can be coupled to the electronic ones via photon-momentum
exchange. The laser excitation can be done in several different ways, giving
rise to a large number of possible interaction Hamiltonians. Here we will be interested in a situation
where the motional sidebands are well resolved and the so-called Lamb--Dicke
limit applies. Moreover, we will only consider the excitation of one
collective motional degree of freedom, the center of mass (CM) motional mode in the longitudinal trap direction.

In order to perform the operations requested by the protocol, we have
to consider the laser excitation of any given electronic transition of an ion
in three different ways. One consists in illuminating the ion with laser light resonant with the transition, often called {\em carrier} excitation. The second way is to excite the transition with light resonant with the first lower motional sideband (red sideband); and the last one uses light resonant with the first higher sideband (blue sideband).

Under the conditions stated above, the interaction Hamiltonians corresponding to each one of these situations are given, in the interaction picture, by~\cite{vogel:4214}:
\begin{eqnarray}
\hat{H}_{\rm C}&=&\frac{1}{2}\hbar\,|\Omega|\,e^{i\phi}\,\hat{\sigma}_+ +\mbox{h.c.}\label{eqham0}\\
\hat{H}_{\rm R}&=&\frac{i}{2} \eta \hbar\,|\Omega|\, e^{i\phi}\, \hat{\sigma}_+\,
\hat{a}+\mbox{h.c.}\label{eqham-1}\\
\hat{H}_{\rm B}&=&\frac{i}{2} \eta \hbar\,|\Omega|\, e^{i\phi}\, \hat{\sigma}_+\,
\hat{a}^\dagger+\mbox{h.c.}\label{eqham+1},
\end{eqnarray}
respectively.  Here the operator $\hat{\sigma}_+$ is the electronic raising
operator, and $\hat{a}$ and $\hat{a}^{\dagger}$ are the annihilation and
creation operators of the CM vibrational mode,
respectively. $\Omega=|\Omega|e^{i\phi}$ is the laser Rabi frequency and
$\eta$ is the Lamb--Dicke parameter, which in the Lamb--Dicke limit satisfies $\eta\ll 1$.

The rotations appearing in the protocol are performed via carrier excitation of the ions, according to the time evolution operator:
\begin{equation}
e^{-\frac{i}{\hbar}\hat{H}_{\rm C}\tau}=\hat{U}_{\rm C}(\theta,\phi)=e^{-i\frac{\theta}{2}\left(\cos(\phi)\,\hat{\sigma}^{ge}_x -\sin(\phi)\,\hat{\sigma}^{ge}_y\right)}\;,
\label{evolham0}
\end{equation}
where $\theta=|\Omega|\tau$ and $\hat{\sigma}^{ge}_\alpha$ are the electronic Pauli operators
(see Eq.~\eqref{subspace sigmas}), acting on the subspace spanned by $\ket{g}$ and $\ket{e}$, these in turn representing generic ground ($\ket{G}$, $\ket{G'}$, $\ket{g_q}$) and excited ($\ket{E}$, $\ket{E'}$, $\ket{e_q}$) states, respectively. In particular,
the rotations $\hat{R}^{ge}_x(\theta)=\hat{U}_{\rm C}(\theta,0)$ and
$\hat{R}^{ge}_y(\theta)=\hat{U}_{\rm C}(\theta,-\pi/2)$ are obtained by adjusting the
laser phase. 

Controlled operations are achieved through excitation of the CM vibrational mode. Both Jaynes--Cummings (red sideband, Eq.~\eqref{eqham-1}) and anti Jaynes--Cummings (blue sideband, Eq.~\eqref{eqham+1}) interactions are used in the protocol.

For concreteness, we present an implementation using $^{40}{\rm Ca}^{+}$ ions, which have been
used in several experiments in Innsbruck~\cite{riebe:734}. States $\ket{g_q}$
and $\ket{e_q}$ of each qubit can be encoded in sublevels $m=-1/2$ of the
$4S_{1/2}$ state and $m=-1/2$ of the $3D_{5/2}$ state, respectively.

\begin{figure}[htbp]
\resizebox{!}{4cm}{\includegraphics{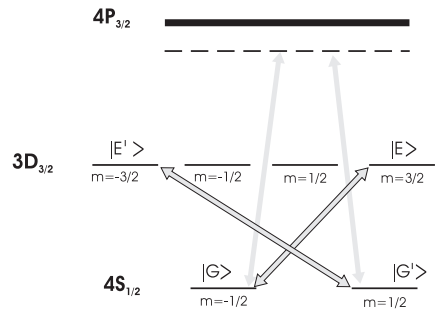}}
\caption{Encoding of the ancillary ion and relevant transitions} 
\label{figure:Levels}\centering
\end{figure}

We further use another $^{40}{\rm Ca}^{+}$ ion for our four level ancilla. States $\ket{G}$ and $\ket{G'}$ can be associated to the $m=-1/2$ and the $m=1/2$ sublevels of the
$4S_{1/2}$ state, respectively, while states $\ket{E}$ and $\ket{E'}$ can be associated to the $m=3/2$ and $m=-3/2$ sublevels of the metastable $3D_{3/2}$ state (see fig.~\ref{figure:Levels}).

The first part of the protocol consists in preparing the ancilla in the
superposition given by Eq.~\eqref{preparedancilla}.  Since
$4S_{1/2}\leftrightarrow3D_{3/2}$ is a quadrupole transition, it is possible
to select sublevel transitions with $\Delta m=\pm 2$ by controlling the angle
between the incident laser beam and the direction of a weak applied magnetic
field, as well as the laser polarization. These transitions allow us to
perform rotations between specific ground and excited levels without
disturbing the rest.  Rotations $R_y$ among the two ground state levels
$\ket{G}$ and $\ket{G'}$ are performed through a carrier Raman excitation,
which can be achieved through simultaneous off-resonant excitation of the
dipole transition $4S_{1/2}\leftrightarrow4P_{3/2}$ by two laser beams of
polarizations $\pi$ and $\sigma_+$ focused on the ancillary ion.

We start with the system in state $\ket{G}\ket{\chi}\ket{0}_v$, where
$\ket{0}_v$ is the ground state of the CM vibrational mode. In order to
perform the first rotation $R_y ^{GE}(\theta_1)$ we apply a carrier excitation
of the $4S_{1/2}\leftrightarrow3D_{3/2}$ quadrupole transition with $\Delta
m=2$. This rotation is followed by a carrier Raman excitation among levels
$\ket{G}$ and $\ket{G'}$ as described above, implementing $R_y
^{GG'}(\theta_2)$. We finally perform the $R_y ^{G'E'}(\theta_3)$ rotation
applying a carrier excitation of the $4S_{1/2}\leftrightarrow3D_{3/2}$
transition with $\Delta m=-2$. After that series of laser pulses, the
electronic state of the ions is given by Eq.~\eqref{preparedancilla}, with the
vibrational CM mode still unoccupied.

Next in the protocol are the controlled
operations. First, we want to apply $\hat\sigma_y$ on $\ket{\chi}$ controlled
by state $\ket{G'}$. We apply on the ancilla a $\pi$-pulse resonant with the first
blue sideband of the $4S_{1/2}\leftrightarrow3D_{3/2}$ transition with $\Delta
m=-2$, and with phase $\phi=0$. (This choice of phase will be maintained for
all red and blue detuned pulses). This anti Jaynes--Cummings interaction
takes state  $\ket{G'}\ket{\chi}\ket{0}_v$ to
$\ket{E'}\ket{\chi}\ket{1}_v$, leaving the other states unchanged. With this
operation we transfer the information about the occupation of state $\ket{G'}$
to the motional state $\ket{1}_v$. We now apply $\hat\sigma_y$ on the qubit, controlled by the motional state
$\ket{1}_v$. We achieve this by performing a carrier rotation
$R^{g_qe_q}_x(-\pi/2)$ on the qubit, followed by a $2\pi$-pulse
resonant with the first red sideband of the transition between level
$\ket{g_q}$ and an auxiliary level $\ket{e'_q}$, which takes
$\ket{g_q}\ket{\chi}\ket{1}_v$ back to itself through
$\ket{e'_q}\ket{\chi}\ket{0}_v$, while gaining a minus sign. We then apply a carrier rotation $R^{g_qe_q}_x(\pi/2)$ on the qubit to obtain state
$-\ket{E'}\hat{\sigma}_y\ket{\chi}\ket{1}_v$. Finally, another $\pi$-pulse on
the ancilla resonant with the first blue sideband of the
$4S_{1/2}\leftrightarrow3D_{3/2}$ transition will bring the state
$-\ket{E'}\hat{\sigma}_y\ket{\chi}\ket{1}_v$ to $\ket{G'}\hat{\sigma}_y\ket{\chi}\ket{0}_v$. Notice that all other states remain unaffected by these transformations.

An analogous procedure is followed in order to act on the qubit with
$\hat{\sigma}_x$, controlled by state $\ket{E}$: we first apply
on the ancilla a $\pi$-pulse resonant with the first red sideband of the
$4S_{1/2}\leftrightarrow3D_{3/2}$ transition with $\Delta m=2$, which
transforms state $\ket{E}\ket{\chi}\ket{0}_v$ into
$-\ket{G}\ket{\chi}\ket{1}_v$. Then we apply on the target ion a
$R^{g_qe_q}_y(\pi/2)$ rotation, followed by a $2\pi$-pulse resonant with the first red
sideband of the $\ket{g_q}\leftrightarrow\ket{e'_q}$ transition, and a
$R^{g_qe_q}_y(-\pi/2)$ rotation. Finally, we apply on the ancilla another
red sideband $\pi$-pulse, identical to the first one.

The $-\hat{\sigma}_z$ operation, controlled by level $\ket{E'}$ can be
achieved by applying the following sequence of pulses: a $\pi$-pulse with
$\Delta m=-2$ resonant with the first red
sideband of the $4S_{1/2}\leftrightarrow3D_{3/2}$ transition on the ancilla, a
$2\pi$-pulse resonant with the first red sideband of the
$\ket{g_q}\leftrightarrow\ket{e'_q}$ transition on the target ion, and another
$\pi$-pulse on the ancilla, identical to the first. After these three
controlled operations are performed, the system is left in the state given by
equation~\eqref{aftercontrol}, with no excitations in the vibrational mode.

We complete the protocol by performing the four rotations: $\hat{R}_{y}^{GE}\left(\frac{\pi}{2}\right)$,
$\hat{R}_{y}^{G'E'}\left(\frac{\pi}{2}\right)$,  
$\hat{R}_{y}^{GG'}\left(\frac{\pi}{2}\right)$ and
$\hat{R}_{y}^{EE'}\left(\frac{\pi}{2}\right)$, by applying successive carrier
laser pulses. We obtain thus the state given by Eq.~\eqref{finalstate}.

At this point, we need to measure the populations of the ancillary electronic
states. This is accomplished, via electronic shelving technique, by exciting
the $4S_{1/2}\leftrightarrow4P_{1/2}$ transition and monitoring the fluorescence
light~\cite{leibfried:281}. In our case, a preliminary series of
laser pulses is needed in order to prepare the ancillary ion for measurement.

First, a carrier Raman $\pi$-pulse excites the
$3D_{3/2}\leftrightarrow3D_{5/2}$ transition via the $4P_{3/2}$ level, using two
$\pi$-polarized laser beams. This brings the population of state $\ket{E}$
to the Zeeman sublevel $3D_{5/2}$ $(m=3/2)$ and the population of $\ket{E'}$ to
$3D_{5/2}$ $(m=-3/2)$. Then, a $\pi$-pulse with $\Delta m=2$ resonant with the
$4S_{1/2}\leftrightarrow3D_{3/2}$ transition is applied, bringing the
population of state $\ket{G}$ to $3D_{3/2}$ $(m=3/2)$, followed by a $\pi$-pulse with
$\Delta m=0$ resonant with the $4S_{1/2}\leftrightarrow3D_{5/2}$ transition,
bringing the population of state $\ket{G^\prime}$ to $3D_{5/2}$ $(m=1/2)$.
Finally, a $\pi$-pulse with $\Delta m=2$ resonant with the
$4S_{1/2}\leftrightarrow3D_{3/2}$ transition is applied, bringing the population of state $\ket{G}$
back to level $4S_{1/2}$ $(m=-1/2)$.

Now transition $4S_{1/2}\leftrightarrow4P_{1/2}$ is excited and fluorescence light is
monitored. If fluorescence is observed, this indicates that the ancillary
state $\ket{G}$ was occupied, and the measurement ends. Otherwise, a
$\pi$-pulse with $\Delta m=0$ resonant with the
$4S_{1/2}\leftrightarrow3D_{5/2}$ transition is applied, bringing the population of state
$\ket{G'}$ back to level $4S_{1/2}$ $(m=1/2)$. The fluorescence test is
repeated, a positive result indicating that state $\ket{G'}$ was
occupied. Again, if no fluorescence is observed, a $\pi$-pulse with $\Delta
m=2$ resonant with the $4S_{1/2}\leftrightarrow3D_{5/2}$ transition is applied, bringing the
population of state $\ket{E}$ to $4S_{1/2}$ $(m=-1/2)$. The fluorescence test
is repeated once more, now a positive result indicating occupation of the
$\ket{E}$ state. No light observed in this last test indicates that state
$\ket{E'}$ was the one occupied. Several iterations of this process yield the
occupation probabilities of the four ancillary levels. This whole procedure is
equivalent to measuring the probability distribution of the spectrum of a single observable: the electronic
energy of the ancilla.

We wish to note that the whole series of pulses would be considerably reduced
by taking advantage of the Zeeman splitting of the levels obtained with a
strong applied magnetic field, but this would also limit the initial
Doppler cooling of the ions~\cite{thesis}.

We would like to stress that our protocol is designed for pure
states. However, if we consider small deviations from a pure state we still
may have a good estimate for the concurrence. Assuming, for example, a density
matrix of the form $\rho=\lambda \rho^\prime+(1-\lambda)\ket{\chi}\bra{\chi}$, where $\rho^\prime$, is a separable state
and $\lambda\ll 1$, then one can show that the difference between the values
of the square of the true concurrence and the value obtained using the above
protocol is $-2\lambda(1- \vec{P}\cdot\vec{P^\prime}) +{\cal O}(\lambda^2)$,
where $\vec{P^\prime}$ and $\vec{P}$ are the Bloch vectors associated with the
reduced density matrices of $\rho'$ and $\ket{\chi}\bra{\chi}$, respectively.

%
%

Before concluding we would like to briefly compare our
protocol with measurements involving simultaneous copies~\cite{florian,walborn06}. Clearly,
the main difference lies in the fact that we do not require simultaneous copies
of the state in order to measure concurrence. Another advantage of our
protocol is that, given that the relation stated in Eq.~\eqref{concu} holds
for the bipartite concurrence of a qubit with an arbitrary number of other qubits,
provided they are all in a pure state, our protocol is trivially generalized
to this case. It is also important to note that we do not have an extra
cost for not using simultaneous copies: the four dimensions of our ancilla match
those in the two qubits where the copy of the system is provided.

As compared to standard tomography, our protocol not only involves a single observable, but is also minimal and optimal, implying in particular that improved accuracy is achieved in determining the state of the system. 

In summary, we presented a minimal and optimal tomographic protocol which
involves the measurement of a single observable. This in turn is used to obtain the concurrence of a pure two qubit state. We also propose a realistic implementation of the protocol for a system of trapped ions, which could in principle be carried out with present day experimental techniques.

%
%

\begin{acknowledgments}
  We would like to acknowledge fruitful discussions with C. Saavedra
  , S. Walborn, L. Davidovich and P.H. Souto Ribeiro. This work was supported by the Brazilian
  agencies CAPES, CNPq, FAPERJ, FUJB, and the Millennium Institute for Quantum
  Information and the Fondecyt 7060168, 1030189 and Milenio ICM P02-49 projects.
\end{acknowledgments}

%
%

\end{document}